\documentclass[apjl]{emulateapj}
\newcommand{\beq}{\begin{equation}}
\newcommand{\beqa}{\begin{eqnarray}}
\newcommand{\eeq}{\end{equation}}
\newcommand{\eeqa}{\end{eqnarray}}
   
\newcommand{\etal}{{\it et al. }}

\newcommand{\gsim}{\ga}

\shorttitle{Deci hertz Laser Interferometer}
\shortauthors{Takahashi \& Nakamura}
\begin{document}

\title{ Deci hertz Laser Interferometer
  can determine the position
of the Coalescing Binary Neutron Stars within an arc minute
a week before  the final merging  event to Black Hole}
\author{Ryuichi Takahashi and Takashi Nakamura}
\affil{
Department of Physics, Kyoto University,
Kyoto 606-8502, Japan 
}


\begin{abstract}
 It may be possible to construct a laser interferometer
gravitational wave antenna in space with
 $h_{rms}\sim 10^{-23}$ at $ f\sim 0.1{\rm Hz}$
in $\sim 2020$.  This deci hertz antenna may be called
 DECIGO/BBO, which stand for DECi hertz Interferometer
Gravitational wave Observatory and Big Bang Observer,
respectively. 
  The  analysis of 1-10 years observational data
 of the coalescing binary neutron stars or black holes
at the distance of $\sim$ 300Mpc will give us the spatial position
 within $\sim$ an arc minute and the time of the coalescence within
 $\sim 0.1$ sec beforehand.
With the knowledge of the accurate position
 and the time of final merging event, the follow up simultaneous
 observation using
 high frequency ($ f\sim 100{\rm Hz}$) gravitational wave
 antennae as well as electro-magnetic wave
antennae from the radio frequency to the ultra high energy
 gamma ray will reveal the physics in the enigmatic
event of the coalescence and the formation of the black hole.
\end{abstract}



\section{Introduction}

There are at least four frequency bands where active or
 planned gravitational wave antennae exist. 
The resonant bars as well as ground laser interferometers
such as TAMA300
, LIGO I
, VIRGO
, and GEO600
 are covering the frequency band of
 10Hz--kHz while the timing analysis of  pulsars is covering
$\sim 10^{-8}$Hz band. Laser Interferometer Space Antenna
\footnote{LISA web page http://lisa.jpl.nasa.gov/index.html}\
 (LISA) will cover $10^{-4}-10^{-2}$Hz from $\sim$ 2010.
Very recently the space antenna in the deci Hertz band
 ($10^{-2}-10$ Hz) comes into the group.
According to Seto, Kawamura and Nakamura (2001)
it might be possible to construct a laser interferometer
gravitational wave antenna in space with
 $h_{rms}\sim 10^{-27}$ at $ f\sim 0.1{\rm Hz}$
in this century.   Using this antenna they show that 1) $\sim 10^5$
 chirp signals of 
coalescing binary neutron stars per year may be detected with S/N
$\sim 10^4$. 2) The time variation of the Hubble parameter of our
 universe may be determined for ten years observation of binary
 neutron stars at $z\sim 1$ so that we can  directly
  measure  the acceleration of the universe. 3) the stochastic gravitational
 wave which is  predicted by the inflationary universe paradigm,
 may be detected.

Seto \etal (2001) call the deci Hertz antenna 
 DECIGO(DECi hertz Interferometer Gravitational
wave Observatory)  while recently in the NASA
SEU 2003 Roadmap ``Beyond Einstein'' the deci Hertz antenna is
 called as  BBO (Big Bang Observer)
\footnote{See http://universe.gsfc.nasa.gov/be/roadmap.html}\
 to stress the detection of the 
stochastic gravitational waves from inflation.
Since DECIGO and BBO are similar antennae,
in this Letter
 we call the deci hertz antenna as DECIGO/BBO.
In DECIGO,  $h_{rms} \sim 10^{-27}$ around 0.1 Hz is the ultimate goal
 in the far future, assuming the quantum
limit sensitivity for a 100 kg mass.

Let us consider more realistic parameters which might be
achieved in $\sim 2020$ such as 300W  laser power,
3.5m mirrors, $5\times 10^4$km arm length and 0.01 LISA
 acceleration noise, which  will give $h_{rms} \sim 10^{-23}$
 around 0.1 Hz (e.g. Larson, Hiscock \& Hellings 2000). 
In this sensitivity, S/N of the chirp signal from the coalescing
 binary neutron stars  at $z\sim 1$
will be $\sim 1$ while that of 10$M_\sun$ black hole binary
 will be $\sim 10$ so that we may directly measure the
acceleration of the universe similar to the ultimate DECIGO.
After subtracting the signals from neutron star binaries and 
 black hole binaries
\footnote{There may be other sources that we do not
 know about at this point.}\
we may detect the stochastic gravitational waves from inflation 
if  $\Omega_{GW}\gsim 10^{-15}$
which is  the upper bound from  CMB quadrupole anisotropies
 measured by COBE (Gorski {\it et al.} 1996).

One of the main targets of the ground laser interferometer
gravitational wave
antennae is the chirp signal from the coalescing binary
neutron stars or black holes at the distance of $200-400$Mpc
 with the expected event rate of $\sim $1 per year
 (e.g. Phinney 1991).
Since the event rate is $\sim $1 per year, the same binary
is in the band of DECIGO/BBO, $1\sim 10$ years before the
ground laser interferometers and bar detectors
detect the gravitational waves from the final merging event.
This means that in practice we may observe the evolution
of the coalescing binary from $10^{-2}$ Hz band to 
$\sim$kHz band. The duration in the
band of the ground laser interferometer is three minutes or so
while  the duration in DECIGO/BBO band is 1-10 years so that
we may extract the information of the binary from
the  DECIGO/BBO band observation first. The point here is
that we may know the accurate  position
and the distance to the binary beforehand so that we can prepare
 various detectors for the observation of the final merging phase,
 that is, we can point electro-magnetic antennae from radio to
 ultra high energy gamma rays and cosmic rays to the direction of the
coalescing binary at the expected final merging  time.
We may also tune the gravitational wave antennae to
various sensitivity bands to detect the characteristics of
 the merging event such as the frequency at the inner most
 stable circular orbit and the quasi-normal
mode of the final black hole.

In this Letter  using the practical DECIGO/BBO
in $\sim 2020$ we show how accurately we can determine 
the spatial position and the distance to the coalescing
binary neutron stars or black holes.
Since the angular resolution of LISA was calculated in
 (Cutler 1998; Hughes 2002; Seto 2002; Vecchio 2003), 
we apply these methods to the case of DECIGO/BBO.
We adopt the Hubble
 parameter $h\equiv H_0$/100km/s/
Mpc=0.7  and the units of $c=G=1$. 

\section{Gravitational Waveform and Parameter Estimation}

\subsection{Sensitivity of DECIGO/BBO}
DECIGO/BBO would consist of three spacecrafts separated by
 $5 \times 10^4$ km which is 1/100 times smaller than 
the size of LISA (see Bender \etal 2000).
They would be spaced in an equilateral triangle and be
orbiting around the Sun.
The change of the detector's orientation and the position
   enables us to obtain  the source position
  in the following two ways;
(i) DECIGO/BBO's orientation rotates with a period of one-year,
 which imposes the modulation on the measured signal.
(ii) DECIGO/BBO's orbital motion around the Sun imposes
 the periodic Doppler shift on the signal frequency
 (see also Cutler 1998 in the case of LISA). 
For the monochromatic signal of the frequency $f$,
 (i) the rotational modulation changes the frequency $f$ to $f \pm 2/T$,
 where $T=1$ yr ($2/T \sim 10^{-7}$ Hz),
 (ii) the Doppler modulation changes $f$ to $f(1 \pm v)$,
 where $v \sim 10^{-4}$.
Then for the frequency higher than $\sim 10^{-3}$ Hz
like in DECIGO/BBO,  the effect of
 the Doppler modulation is more important for the determination of the
 source position than the rotational modulation so that we only
 consider the former effect in the measured signal.
The observed waveform $h$ in the frequency domain
 $\tilde{h}_D(f)$ is given by,
\beq
 \tilde{h}_D(f)=\tilde{h}(f) ~e^{i \phi_{D} (f)}
\label{hcd}
\eeq
where $\tilde{h}(f)$ is the waveform at the solar system barycenter.
 $\phi_D(f)$ is the Doppler phase, which is the difference of the phase
 between the detector and the Sun; 
 $\phi_D=2 \pi f R \sin \theta_S
 \cos (2 \pi t/T-\phi_S)$, where $R=1$ AU, $T=1$ yr and 
 $(\theta_S,\phi_S)$ is the direction to the source.
These angular coordinates are defined in a fixed barycenter
 frame of the solar system (see Cutler 1998).

The strain sensitivity of DECIGO/BBO is about 1000 times better than that of
 LISA (i.e. $10^{-23} ~\mbox{Hz}^{-1/2}$ at $f=0.1-1$ Hz),
and the acceleration noise is 100 times lower than that of
 LISA (i.e. $3 \times 10^{-17} \mbox{m} / \mbox{s}^2 /
 \mbox{Hz}^{1/2}$).
Adopting these parameters, we show the sensitivity of
 DECIGO/BBO as compared with LISA and LIGO II in Fig.1
 (Larson \etal 2000). Note that
we are using a smooth curve and neglecting the wavy
 behavior of the transfer function for simplicity.

\subsection{Gravitational Waveform}
We consider the equal mass neutron star
 and black hole
 binaries as sources of DECIGO/BBO.
We only consider  a circular orbit since the
expected eccentricity is $\sim 10^{-3}$ for the neutron star binary.
We use the restricted 1.5 post-Newtonian approximation 
 as the in-spiral waveform (Cutler \& Flanagan 1994) for simplicity.
The waveform in the frequency domain is given by,
\beq
 \tilde{h}(f) = \mathcal{A} f^{-7/6} e^{i \Psi(f)},
\label{hc}
\eeq
where $\mathcal{A}$ is the amplitude and  $\Psi(f)$
is the phase. They depend on six
 parameters; the red-shifted chirp mass
 $\mathcal{M}_z = (M_1 M_2)^{3/5} (M_1 + M_2)^{-1/5} (1+z_S)$,
 the reduced mass $\mu_z=M_1 M_2 (1+z_S)/(M_1+M_2)$,
 the spin-orbit coupling constant $\beta$,
the coalescence time $t_c$ and
the phase $\phi_c$ and
 the luminosity distance to the source $D_S$.
The amplitude is given by $\mathcal{A}= K \sqrt{5/96} 
 ~\mathcal{M}_z^{5/6} / (\pi^{2/3}  D_S)$, where
$K$ is the constant determined by the inclination of the source, 
 the relative orientation of the source and
 the detector.
Since  the average value of $K$ is about unity (Finn \& Chernoff 1993),
 we assume $K=1$ for the following calculation. 
The phase $\Psi(f)$ in Eq.(\ref{hc})
 is a rather complicated function of $\mathcal{M}_z$,
 $\mu_z$, $\beta$, $\phi_c$ and $t_c$ which is given in Eq.(3.24)
 of Cutler \& Flanagan (1994).

\subsection{Parameter Estimation}

The signal observed by DECIGO/BBO, $\tilde{h}_D (f)$, can be
 obtained inserting Eq.(\ref{hc}) to  Eq.(\ref{hcd}).
The signal $\tilde{h}_D (f)$ is characterized by eight parameters 
 ($\mathcal{M}_z$, $\mu_z$, $\beta$, $\phi_c$, $t_c$, $D_S$,
 ${\theta_S}$, ${\phi_S}$).
In the matched filter analysis with the template, these parameters
 can be determined. 
We compute the errors in the estimation of these parameters using the
 Fisher matrix formalism (Finn 1992; Cutler \& Flanagan 1994).
The variance-covariance matrix of the
 parameter estimation error $\Delta \gamma_i$ is given
 by the inverse of the
 Fisher information matrix $\Gamma_{ij}$ as
 $\langle \Delta  \gamma_i \Delta \gamma_j \rangle
 = \left( \Gamma^{-1} \right)_{ij}$.
The Fisher matrix becomes
\beq 
  \Gamma _{ij}
 = 4 {\mbox{Re}} \int \frac{df}{Sn(f)}~
 \frac{\partial \tilde{h}_{D}^{*}(f)}{\partial \gamma_i}
 \frac{\partial \tilde{h}_{D}(f)}{\partial \gamma_j},
\label{fis}
\eeq
where $Sn(f)$ is the noise spectrum.
We regard $Sn(f)$ as the instrumental noise in Fig.1,
 neglecting the binary confusion noise since there are no
 or little confusion noise for $f \gsim 0.1$ Hz
 (Seto \etal 2001).
The signal to noise ratio ($S/N$) is given by
\beq
  (S/N)^{~2} = 4 \int \frac{df}{Sn(f)}~
 \left| \tilde{h}_D(f) \right|^2.
\label{snr}
\eeq
We integrate the gravitational waveform
 in Eq.(\ref{fis}) and Eq.(\ref{snr}) from 1
or 10 yr before the final merging to the
 cut-off frequency $f_{cut}$ when the binary separation
 becomes $r=6 (M_1 + M_2)$. 

The initial frequency is given by 
 $f_{init}=0.23 (\mathcal{M}_z/M_{\odot})^{-5/8}
 ( T_{obs}/1 \mbox{yr} )^{-3/8} \mbox{Hz}$
where $T_{obs}=1$ or $10$ yr, and the cut-off frequency is
 $f_{cut}=4.4 \times 10^3 [(M_{1 z}+M_{2 z})/M_{\odot}]^{-1} \mbox{Hz}$. 
The result does not depend on the value of $f_{\mbox{cut}}$ so much,
 since $Sn(f)$ is large at $f_{\mbox{cut}}$.

\section{Results}

We consider the neutron star binaries ($1.4 M_{\odot}+1.4 M_{\odot}$)
 and the stellar mass BH binaries ($10 M_{\odot}+10 M_{\odot}$) 
 at $D_S=200 \mbox{h}^{-1}$ Mpc and the intermediate mass BH binaries
 of mass ($10^2 M_{\odot}+10^2 M_{\odot}$)
as well as ($10^3 M_{\odot}+10^3 M_{\odot}$) at
 $D_S=3000 \mbox{h}^{-1}$ Mpc (Hubble distance) as the sources.
Note here that we adopt $h=0.7$.
For each mass case we randomly distribute $10^4$ binaries on the celestial
 sphere at $D_S$,
 and we show the probability distribution of the angular resolution
 for these sources.

In Fig.2, we show the relative probability distribution of the
 angular resolution for the various binary masses 
$M_{1,2z}=1.4 M_{\odot}, 10^{1-3} M_{\odot}$
in the case of $1$ yr observation. 
The solid lines and the dashed lines show $\Delta \theta_S$ and
 $\Delta \phi_S$, respectively.
We also show the signal to noise ratio $S/N$
 and the estimation error
 in the coalescence time $\Delta t_c$ assuming
the template is accurate enough.
 $S/N$ is independent of $\theta_S$ and $\phi_S$
 since the phase factor $e^{i \phi_{D} (f)}$
  becomes unity in Eq.(\ref{snr}) while  $\Delta t_c$
  is found to depend on $\theta_S$ and $\phi_S$ 
 very weakly. In general, 
the errors ($\Delta \theta_S, \Delta \phi_S, \Delta t_c$) simply scale
 as $(S/N)^{-1}$.
As shown in Fig.2, the angular resolutions are typically $\sim 0.1-10$
 arc minutes for the neutron star binaries and the
 black hole binaries.
This is about $10-1000$ times better than that of LISA ($\sim 1$ deg).
 The Doppler modulation ($\phi_D \propto f$ in
 Eq.(\ref{hcd})) effect is larger than the rotational
modulation for the high frequency band like DECIGO/BBO.
It is shown  (Cutler \& Vecchio 1998;
 Moore \& Hellings 2002; Takahashi \& Seto 2002),
that the angular resolution scales as $\propto f^{-1} (S/N)^{-1}$
 for the monochromatic sources.
Thus the angular resolution of DECIGO/BBO should be
 about 100 times better
 than that of LISA if the S/N is equal.
This explains the accuracy of  $\sim 0.1-10$ arc minute.

We comment the results for $10$ yr observation.
In this case, the angular resolution is $2-4$ times better than the case
 for $1$ yr observation in Fig.2.

In Table.1,
we show the $S/N$ and the estimation errors of the chirp mass
 $\mathcal{M}_z$, the reduced mass $\mu_z$, the coalescence time
 $t_c$ and the distance to the source $D_S$.  
The results are presented for the various binary masses
 ($1.4,10^{1-3} M_{\odot}$) for 1yr (the upper line)
  and 10 yr (the lower line) observation.
The errors simply scale as $(S/N)^{-1}$ for the
change of the distance to the source.
For the accuracy of the chirp mass $\Delta \mathcal{M}_z \sim 10^{-7}$
 and the reduced mass $\Delta \mu_z \sim 10^{-4}$,
 each mass of the binary is determined within $\sim 10^{-4}$.
The distance to the binary is determined by $\simeq (S/N)^{-1}$.

\section{Discussions}

The distance to gamma ray bursts (GRBs) was not determined
before 1997. The minimum possible distance was $\sim$
100 AU and the maximum one was the Hubble distance (Fishman and Meegan
 1995) so that  many theoretical models
were equally possible.  The main reason why the
distance to GRBs was not determined is that the accuracy
of the spatial position of GRBs using the gamma ray
observation was at most $\sim$ deg so that  many
possible host galaxies exist in an error box of GRBs.
To find the host galaxy of GRB was akin to searching a needle in a
 haystack.  In 1997 Italian-Dutch satellite Beppo-Sax succeeded
 in obtaining the  high resolution ($\sim$ 2 arc minutes)
 X-ray images of GRBs which led to the determination of
 redshifts and host galaxies of GRBs (Costa et al 1997). 
This clearly demonstrates that the accuracy of an arc minute
 or so is indispensable to study the objects in full
 details.

In this Letter we showed that spatial position of
coalescing binary neutron stars or black holes of
mass $\sim 10 M_\odot$  at the distance of 
$\sim 300$ Mpc can be determined within an arc minute
from one year observation. This accuracy is comparable
or better than that of Beppo-SAX so that we may expect
the rapid progress of research like in GRBs after 1997.
\footnote{
The situation for GRBs is a little different from that for
 coalescing compact binaries.
For compact binaries, we can determine the distance without
 identification of a host galaxy.
But there are $\sim 10^4$ galaxies in one square degree, 
and it would be difficult to identify the host galaxy
even if the distance is determined.  
}\
In Beppo-SAX case the accurate position of GRBs was
obtained a few hours after the GRB event. In our case
the observation period is $1-10$ year before the final
merging to black hole so that we would distribute the
alert of the event beforehand.

Let us consider the observation from $1$ yr before the merging
 up to a day before the merging.
In Fig.4, we show the angular resolution $\Delta \phi_S$
 for the neutron star binary ($1.4+1.4 M_{\odot}$)
 at $200 h^{-1}$ Mpc at each time.
We integrate the waveform in Eq.(\ref{fis})
 from $1$ yr before the final merging
 to three months (the dotted line), a month (the short
 dashed line), a week (the long dashed line) and 
a day (the dot-dashed line) before the merging. 
The solid line shows for the case of 
 full one year observation  up to the cut-off frequency.
 We see that the accuracy is $\sim$ 10 arc minutes at
three months before the final merging while $\sim$ 0.4 arc minute at a
 day before so that the accuracy of the position
depends on the time before the final merging.


In conclusion DECIGO/BBO in $\sim$ 2020 can determine the
  angular position 
 of the neutron star binary and the black hole binary
 at $\sim 300$ Mpc within an arc minute before the
final merging to the black hole. In a sense
DECIGO/BBO will truly open gravitational wave
astronomy since the positional accuracy is comparable
to X-ray telescope on Beppo-SAX.

\acknowledgements
We would like to thank T. Tanaka and N. Seto for useful
 comments and discussions.
This work was supported in part by
Grant-in-Aid for Scientific Research
of the Japanese Ministry of Education, Culture, Sports, Science
and Technology,
No.14047212  and No.14204024.



\begin{figure}
  \plotone{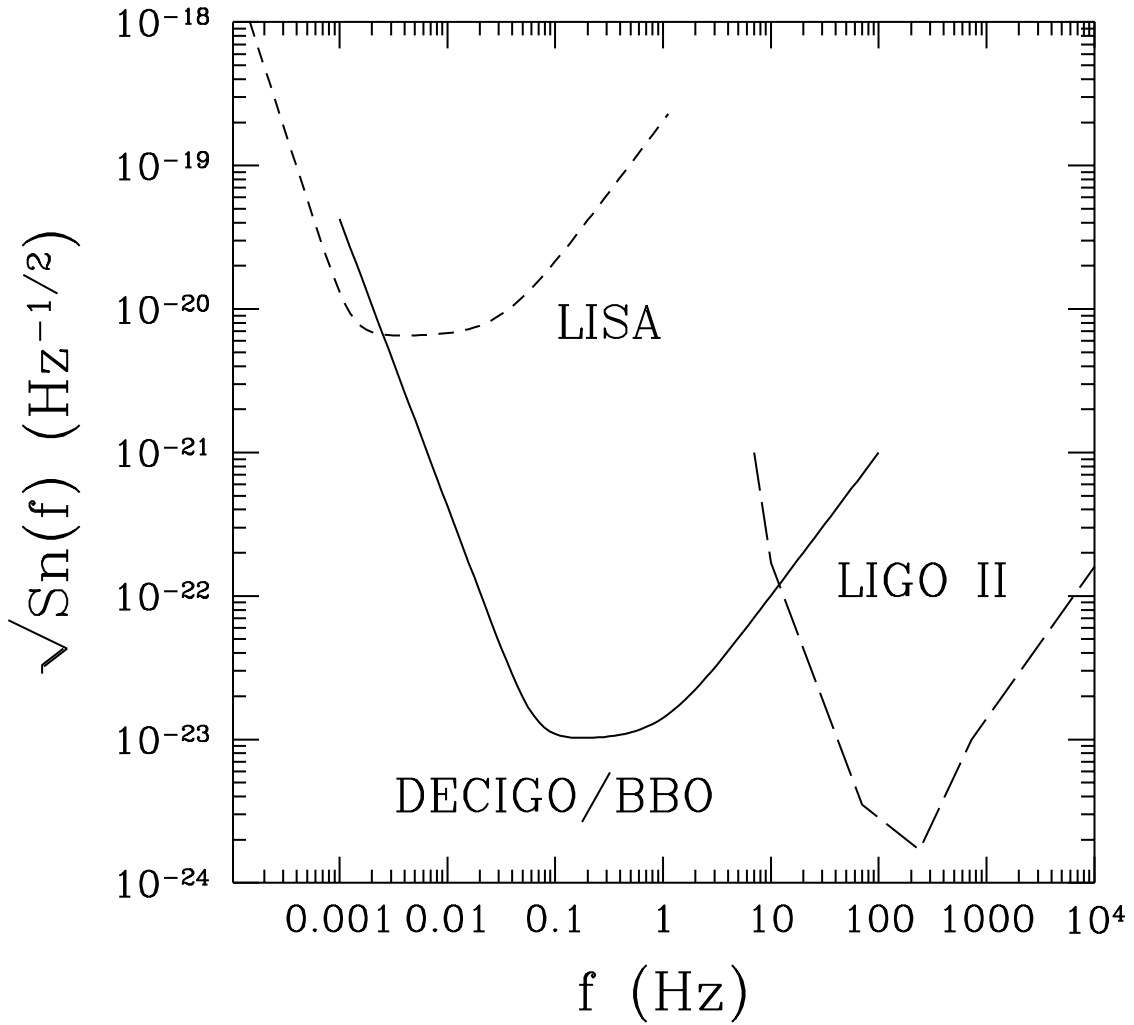}
  \caption{
The strain sensitivity of DECIGO/BBO.
We are using a smooth curve and neglecting the wavy
 behavior of the transfer function for simplicity.}
\end{figure}

\begin{figure}
  \begin{minipage}[t]{7.5cm}
    \vspace{0.1cm}
    \includegraphics[height=7.5cm,clip]{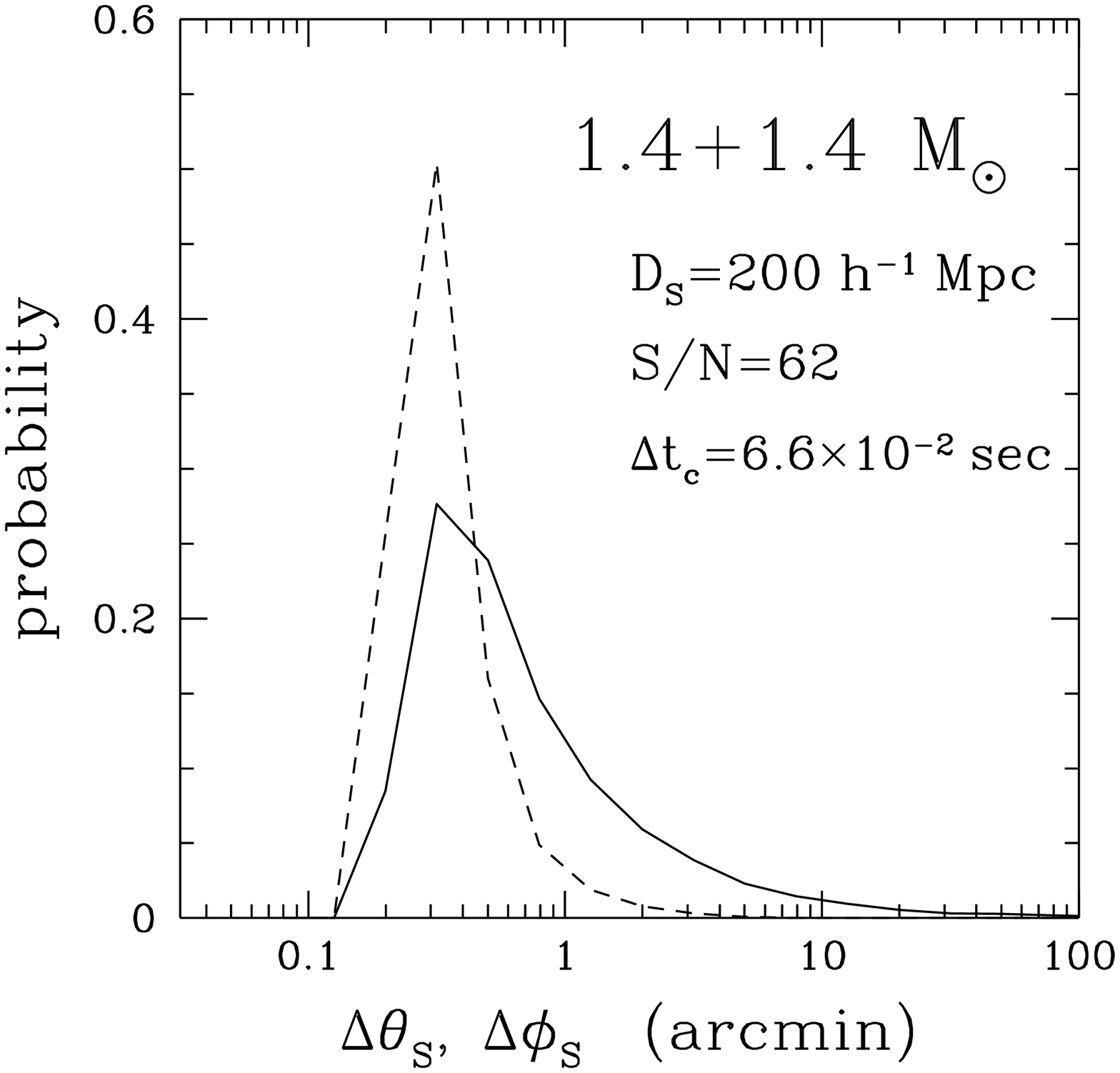}
    \vspace{0.1cm}
  \end{minipage}
  \begin{minipage}[t]{7.5cm}
    \vspace{0.1cm}
    \includegraphics[height=7.5cm,clip]{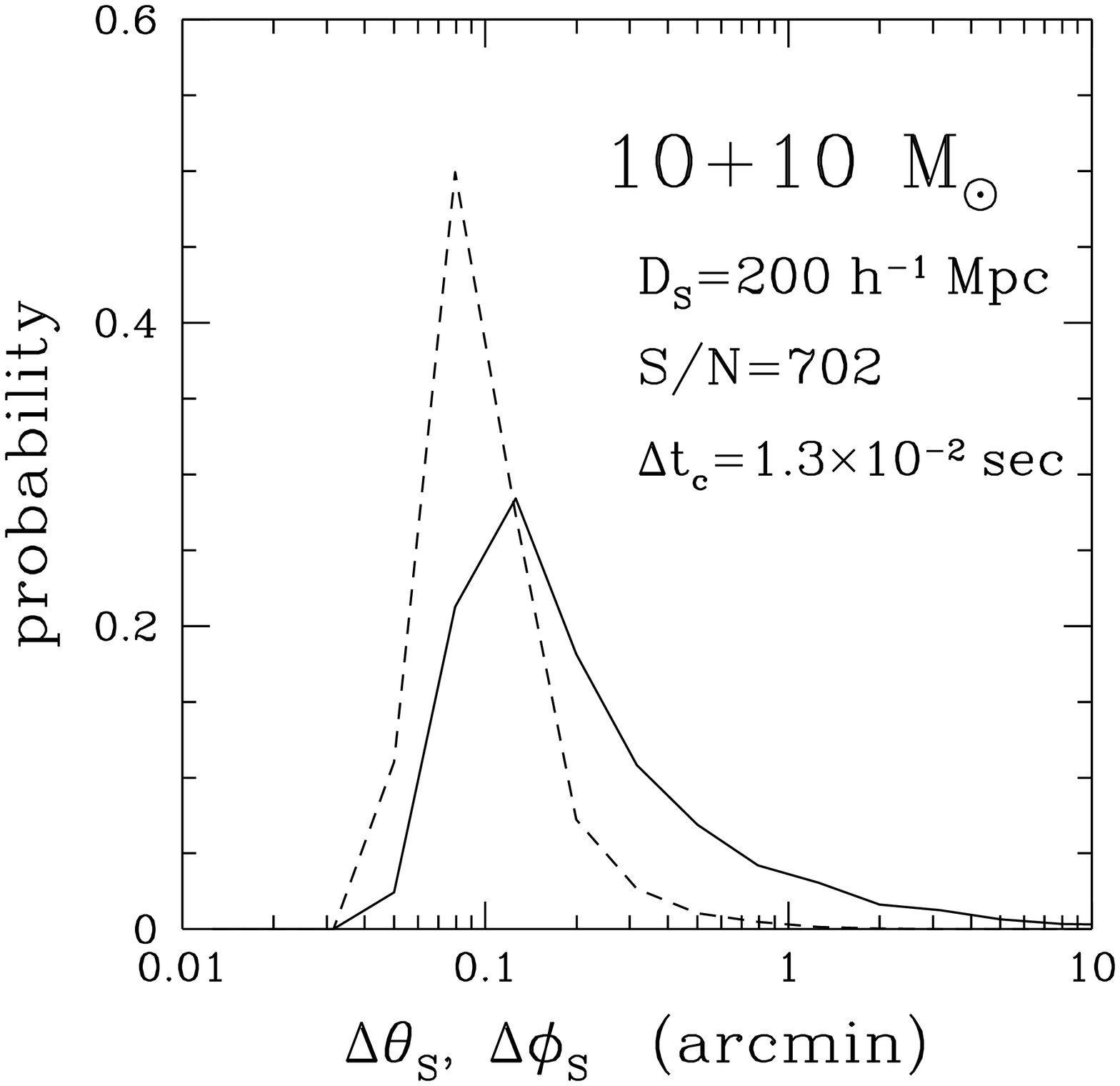}
    \vspace{0.1cm}
  \end{minipage}
    \\
  \begin{minipage}[t]{7.5cm}
    \vspace{0.1cm}
    \includegraphics[height=7.5cm,clip]{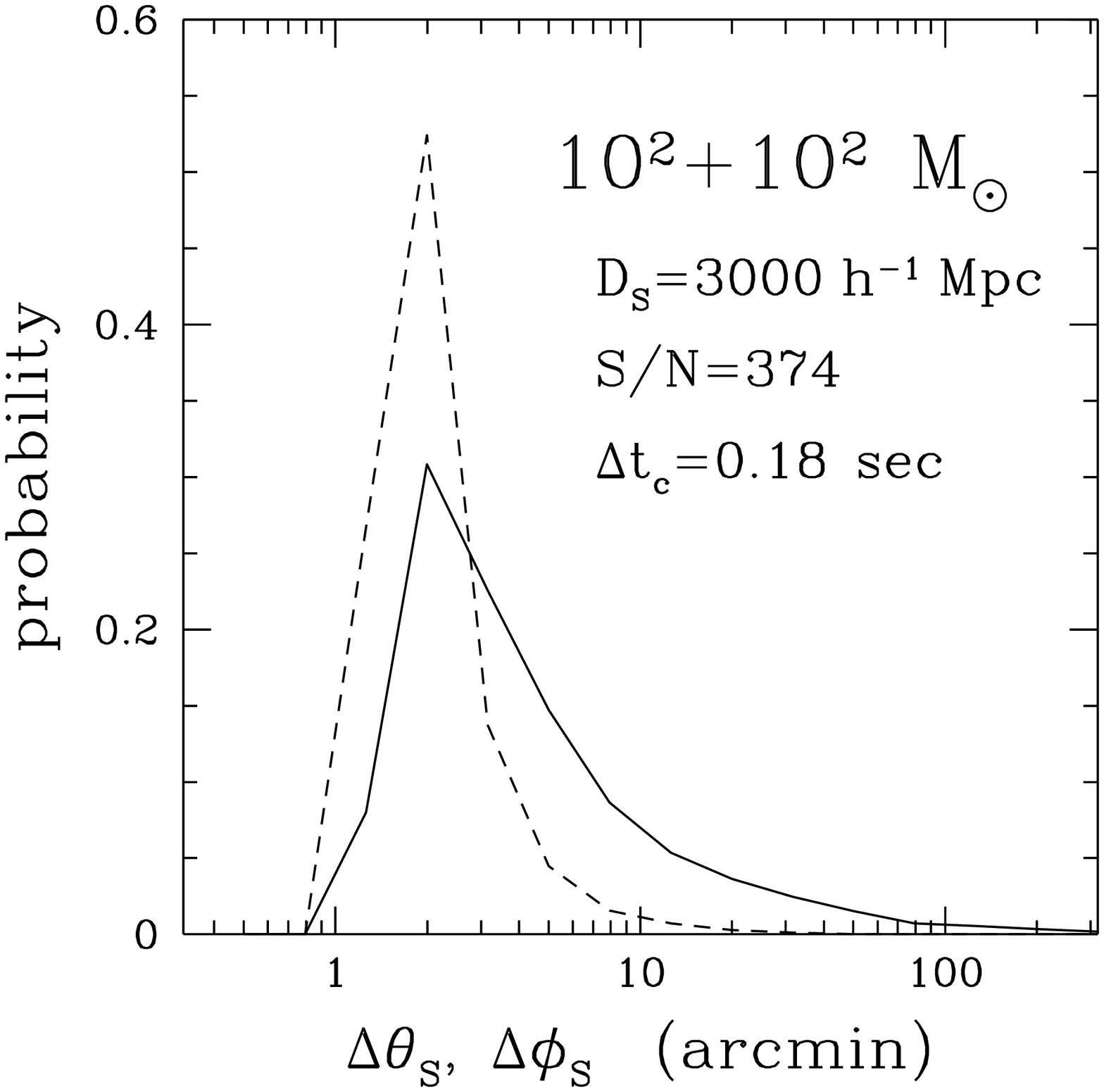}
    \vspace{0.1cm}
  \end{minipage}
  \begin{minipage}[t]{7.5cm}
    \vspace{0.1cm}
    \includegraphics[height=7.5cm,clip]{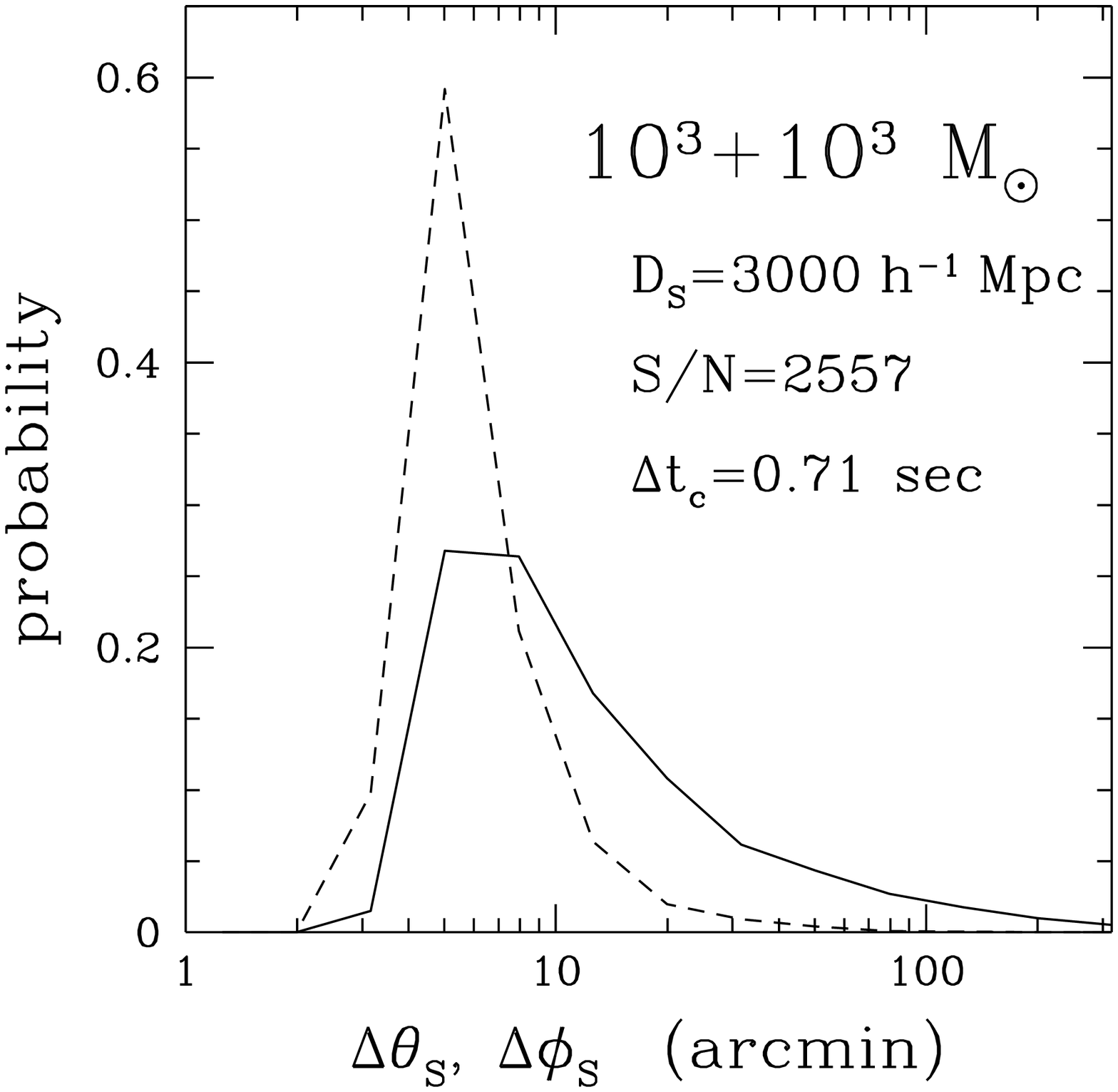}
    \vspace{0.1cm}
  \end{minipage}
    \\
  \caption{
The relative probability distribution of the angular resolution
 of DECIGO/BBO in the case of $1$ yr observation.
For each mass case we randomly distribute
 $10^4$ binaries on the celestial sphere at $D_S$ (the Hubble parameter
 $h=0.7$),
 and we show the probability distribution of the angular resolution
 for these sources. The
 solid lines and the dashed lines show $\Delta \theta_S$ and
 $\Delta \phi_S$, respectively. The signal to noise ratio
 $S/N$ and the estimation error
 in the coalescence time $\Delta t_c$  are also shown.
 $S/N$ is independent of $\theta_S$ and $\phi_S$ 
 since the phase factor $e^{i \phi_{D} (f)}$
  becomes unity in Eq.(\ref{snr}) while  $\Delta t_c$ 
  is found to depend on $\theta_S$ and $\phi_S$
 very weakly.
} 
\end{figure}

\begin{figure}
  \plotone{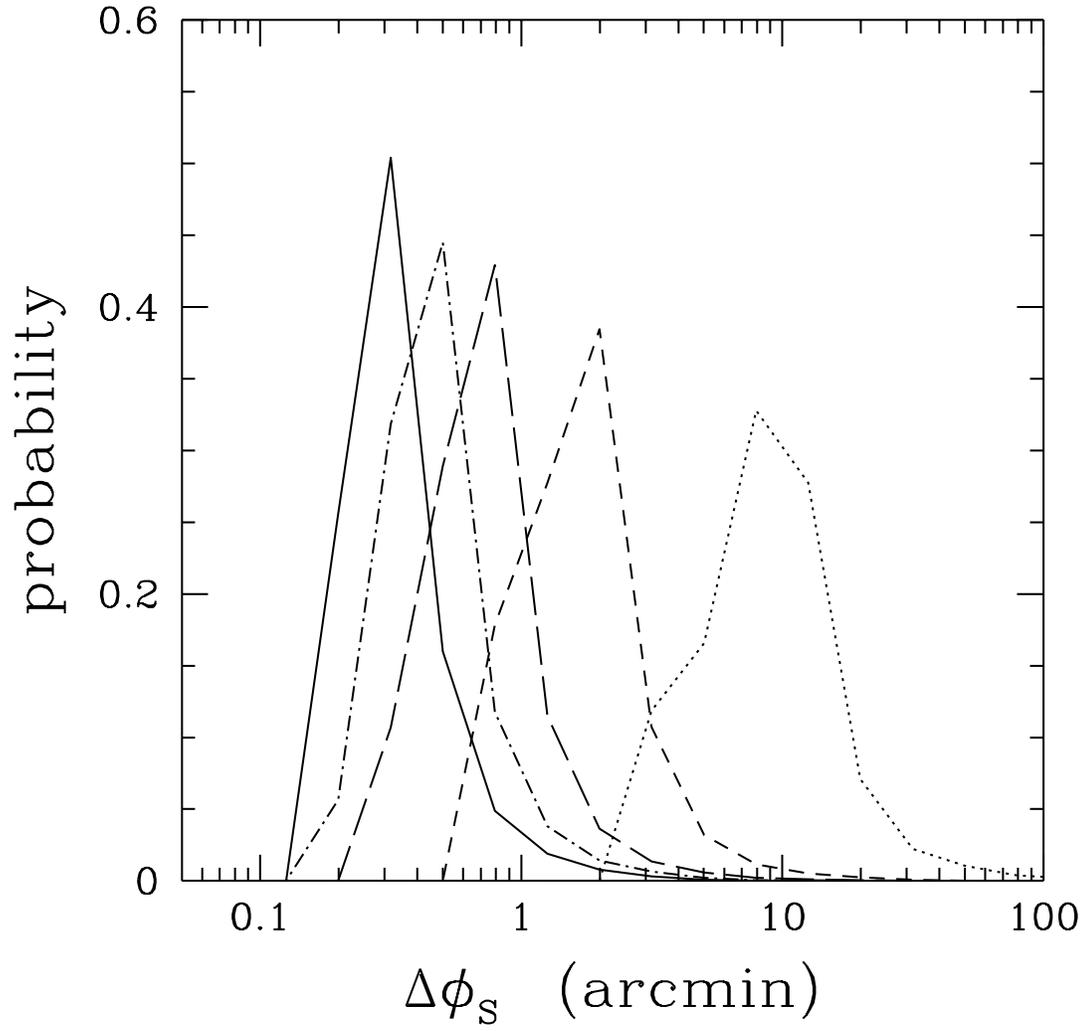}
  \caption{
The angular resolution $\Delta \phi_S$
 for the neutron star binary ($1.4+1.4 M_{\odot}$)
 at $200 h^{-1}$ Mpc ($h=0.7$).
The observational period is
 from $1$ yr before the final merging
 to three months (the dotted line), a month (the short 
 dashed line)  ,  a week (the long dashed line) and 
a day (the dot-dashed line) before the merging. The solid line shows
 for the case of full 
one year observation up to the cut-off frequency.}
\end{figure}

\begin{table}
  \begin{center}
  \setlength{\tabcolsep}{10pt}
  \renewcommand{\arraystretch}{1.1}
  \begin{tabular}{cccccc} \hline\hline
   Binary Masses~($M_{\odot}$) & ~$S/N$~
 & ~$\Delta \mathcal{M}_z$ / $\mathcal{M}_z$~
 & ~$\Delta \mu_z / \mu_z$~ & ~$\Delta t_c$ (sec)~
 & ~$\Delta D_S / D_S$ \\ \hline
   $1.4 + 1.4$~ & ~62~ & ~$2.0 \times 10^{-7}$~ & ~$7.4 \times 10^{-4}$~
 & ~$6.6 \times 10^{-2}$~ & ~$1.6 \times 10^{-2}$~ \\  
 ~ & ~114~ & ~$8.3 \times 10^{-9}$~
 & ~$7.1 \times 10^{-5}$~
 & ~$2.1 \times 10^{-2}$~ & ~$8.8 \times 10^{-3}$~ \\ \hline
   $10 + 10$~ & ~702~ & ~$3.4 \times 10^{-8}$~ & ~$8.4 \times 10^{-5}$~
 & ~$1.3 \times 10^{-2}$~ & ~$1.4 \times 10^{-3}$~ \\     
              ~ & ~812~ & ~$3.5 \times 10^{-9}$~ & ~$1.5 \times 10^{-5}$~
 & ~$4.9 \times 10^{-3}$~ & ~$1.2 \times 10^{-3}$~ \\ \hline
   $10^2 + 10^2$~ & ~374~ & ~$2.7 \times 10^{-7}$~ & ~$2.8 \times 10^{-4}$~
 & ~ 0.18 ~ & ~$2.7 \times 10^{-3}$~ \\
  ~& ~375~ & ~$5.9 \times 10^{-8}$~ & ~$9.3 \times 10^{-5}$~
 & ~ 0.15 ~ & ~$2.7 \times 10^{-3}$~ \\ \hline 
   $10^3 + 10^3$~ & ~2557~ & ~$3.3 \times 10^{-7}$~ & ~$1.4 \times 10^{-4}$~
 & ~ 0.71 ~ & ~$3.9 \times 10^{-4}$~ \\
            ~ & ~2557~ & ~$1.2 \times 10^{-7}$~ & ~$7.5 \times 10^{-5}$~
 & ~ 0.47 ~ & ~$3.9 \times 10^{-4}$~
  \\ \hline\hline
  \end{tabular}
  \end{center} 
\caption{
 $S/N$ and the estimation errors of the chirp mass
 $\mathcal{M}_z$, the reduced mass $\mu_z$, the coalescence
 time $t_c$ and the distance to the source $D_S$. 
The results are presented for the various binary masses
 ($1.4,10^{1-3} M_{\odot}$) for 1yr (the upper line)
 and 10 yr (the lower line) observation.
}
\end{table}

\end{document}